\documentstyle[aps,multicol,epsfig]{revtex}
\newcommand{\nc}{\newcommand}
\nc{\be}{\begin{equation}}
\nc{\ee}{\end{equation}}
\nc{\bea}{\begin{eqnarray}}
\nc{\eea}{\end{eqnarray}}
\nc{\bean}{\begin{eqnarray*}}
\nc{\eean}{\end{eqnarray*}}
\nc{\mb}{\mbox}
\nc{\rnc}{\renewcommand}
\nc{\r}{\mb{\boldmath$r$}}
\nc{\x}{\mb{\boldmath$x$}}
\nc{\A}{\mb{\boldmath$A$}}
\nc{\sa}{\mb{\boldmath$a$}}
\nc{\sss}{\mb{\boldmath$\sigma$}}
\nc{\nab}{\nabla}
\nc{\X}{\sf x}
\newcommand{\Lrule}{
\noindent\vrule width3.45in height.2pt
  depth.2pt \vrule depth0em height1em}
\newcommand{\Rrule}{
\vspace{0.0in}
\hfill
\vrule depth1em height0.pt \vrule
  width3.45in height.2pt depth.2pt
\vspace*{0.01in}
}
\renewcommand{\narrowtext}{\begin{multicols}{2} \global\columnwidth20.5pc}
\renewcommand{\widetext}{\end{multicols} \global\columnwidth42.5pc}

\begin{document}
\draft   

\def\del{\partial}

\title{ Evolution of  $\nu=1$ Bilayer Quantum Hall Ferromagnet
}
\author
{
Kentaro Nomura and Daijiro Yoshioka
 }

\vspace{5mm}

\address{
Department of Basic Science, 
University of Tokyo, 3-8-1 Komaba, Tokyo 153-8902\\
}

\date{\today}
\maketitle

\begin{abstract} 
The natures of the ground state in a $\nu_{\rm T}=1$ bilayer quantum Hall system at 
a variety of layer spacing
are investigated. At small layer separations the system exhibits spontaneous 
interlayer phase coherence.
It is claimed that 
the Halperin's (1,1,1) state is not relevant in the incompressible regime
 near the incompressible to
compressible transition point in which the Josephson-like effect was observed.
The two-particle correlation function shows  the deflated correlation hole
 at this regime.  
An effective model that can give
a good approximation to the ground state 
 is proposed.
A connection to the modified composite fermion theory is discussed.
\end{abstract}


\bigskip
\begin{multicols}{2}

The fractional quantum Hall effect which appears in a two-dimensional
electron system in a strong magnetic field exhibits rich strong correlation 
phenomena.\cite{qhe}
 For a decade, a large number of studies have been made on bilayer quantum
Hall systems which consist of a pair of two-dimensional electron gases 
separated by a distance so small as to be comparable to the
typical spacing between electrons in the same layer.\cite{eis2}
The characteristic parameters of this system are the filling factor 
$\nu_{\rm T}=2\pi l^2n$,
 the distance 
 between the layers $d$, and the interlayer
tunneling rate $\Delta_{\rm SAS}$. Here $n$ is the total electron
 density of the system and $l$ is the magnetic length.
 This system can be mapped to an 
equivalent spin-1/2
system by assigning $\uparrow$($\downarrow$) pseudo spins
to electrons in the upper (lower) layer, 
where the 
actual electron spins are assumed to be polarized.
\noindent
\Lrule\\
In this system, various ground states
and spectacular
 excitations
are realized depending on the parameters.
The Coulomb interaction term consists of direct and exchange terms.
Because the latter  
 term tends to align the pseudo spins of the electrons,
the ground state has a ferromagnetic long range order at $\nu_{\rm T}=1/q$ ($q$ odd)
when $d\simeq 0$.
At finite layer separation, 
the direct part of the
electron-electron interaction produces
a local capacitive charging energy that is minimized
when the two layers have equal electron density and cause the easy-plane(XY) type
symmetry in the pseudospin space.
The ferromagnetic order corresponds to the spontaneous interlayer phase coherence.

The simplest Abelian quantum Hall states for the bilayer
systems at $\nu_{\rm T}=2/(m+n)$ are described by the two-component generalization 
of  the Laughlin wave function first introduced by 
Halperin. \cite{hal}
\end{multicols}
\bea
 \Psi_{m,m,n}= \prod_{i<j}
(z_i^{\uparrow}-z_j^{\uparrow})^m
\prod_{i<j}(z_i^{\downarrow}-z_j^{\downarrow})^{m}
\prod_{i,j}(z_i^{\uparrow}-z_j^{\downarrow})^n 
{\rm e}^{-\frac{1}{4}\sum(|z_i^{\uparrow}|^2+|z_i^{\downarrow}|^2)}.
\eea
\begin{multicols}{2}
\noindent
Based on the numerical diagonalization, 
Yoshioka {\it et al.} investigated that these variational wave functions
give a good approximation to the ground state at a certain $d$ and $\nu_{\rm T}$. \cite{ymg} 
In the case of small layer separation at $\nu_{\rm T}=1$, the ground state
is  $\Psi_{1,1,1}$ state with the ferromagnetic order 
which can also be regarded as an excitonic state.

On the other hand, for an infinite separation, 
the bilayer systems at $\nu_{\rm T}=1$ reduces to two compressible
uncorrelated $\nu=1/2$ systems. This compressible state was discussed by
Halperin {\it et al.} \cite{hlr} as
a Fermi-liquid state of the composite fermions.
\cite{jain} 
The approximated wave function
in this limit may be written as a simple product:
\bea
\Psi_{\rm IFL}=\Psi_{\rm FL}(\{z^{\uparrow}\})\Psi_{\rm FL}(\{z^{\downarrow}\}).
\eea
This state is called the independent Fermi-liquid state.
Here 
 $\Psi_{\rm FL}(\{z\})= P_{\rm LLL}
 {\rm det}M \prod_{i<j} (z_i-z_j)^2\ {\rm e}^{-\frac{1}{4}\sum_i |z_i|^2}$
is the wave function of a Fermi-liquid-like compressible\\
 state at $\nu=1/2$,\cite{rr}
and $P_{\rm LLL}$ is the
  projection operator 
\Rrule\\
to the lowest-Landau-level,
${\rm det}M$ is a Slater determinant with the plane-wave matrix elements 
$M_{ij}={\rm e}^{{\rm i}k_ir_j}$.
Varying the layer separation,
the quantum Hall to no quantum Hall transition occurs.\cite{mac1,murphy}
Based on the Chern-Simons effective field theory, 
several predictions have been proposed theoretically.\cite{wz,ei,moon}
In a recent experimental study, Spielman {\it et al.}\cite{spl} observed resonant
interlayer tunneling at zero-bias in this system.
 They also
have recently observed the quantized Hall drag.\cite{kellogg}
 Because 
the peak of the differential conductance and the quantized Hall drag 
are observed near the transition point,
the properties of the ground state in this regime should be clarified.
Several scenarios have been proposed for the evolution of the ground state as
the layer separation increases from zero to infinity. 
The existence of the charge-density-wave state
with the interlayer coherence near transition point
 is considered in the framework of the Hartree-Fock approximation\cite{cote,brey}
and the effective field theory.\cite{demler}
 Bonesteel {\it et al.} postulate the pairing of the
composite fermions of isolated $\nu=1/2$ layers and obtained a charge gap 
 inversely proportional to $d^2$ which cause the quantum Hall effect.\cite{bonesteel}
 Kim {\it et al.} discussed the paired states
in detail and proposed several candidates of the ground state  wave function.\cite{kim}
In numerical studies,
Schliemann {\it et al.} suggested a single-phase transition
in the pseudospin degrees of freedom,\cite{sgm} and Nakajima found 
that the spectral function
changes qualitatively around transition point.\cite{nakajima}
The estimated critical value 
 agrees with the experiment by Spielman {\it et al.}.
However we do not have a clear understanding of the nature of the ground state
around the transition point.
 In this letter, using the numerical diagonalization on a torus geometry,
we study the details of the evolution of the bilayer system
between small and large $d/l$ regime in the absence of interlayer tunneling.

The kinetic part of the Hamiltonian
 quenches into the lowest Landau level,
so we consider the interaction Hamiltonian
in a projected Hilbert state into the lowest Landau level:
\bea
 H=\frac{1}{\Omega}\sum_{{\bf q}\sigma\sigma'}\frac{1}{2}\
      V_{\sigma\sigma'}({\bf q})\rho_{\sigma}({\bf -q})\rho_{\sigma'}({\bf q}),
\eea
where $\sigma,\ \sigma'$ are pseudospin indices, $\rho_{\sigma}({\bf q})$
is the density operator in the layer $\sigma$, and
$V_{\sigma\sigma'}({\bf q})$ is the Fourier component of 
$V_{\sigma\sigma'}(r)
=e^2/\epsilon {\sqrt {r^2+d^2\delta_{\sigma,-\sigma'}} }$.
When the layer separation $d$ vanishes, 
the Hamiltonian has the pseudospin
SU(2) rotational symmetry that means $[H,{\bf S}]=0$.
  Here ${\bf S}$ is the total pseudospin operator,
\bea
   {\bf S}=\sum_{m=1}^{N} c_{m\sigma}^{\dag}
\frac{\sss_{\sigma\sigma'}}{2} c_{m\sigma'},
\eea
where $m$ means a single particle electron state and $N$ is the number of electron.
 Then the ground state $|\Psi\rangle$
can be also the eigen state of the ${\bf S}^2$ and $S_z$;
\bea
 {\bf S}^2|\Psi\rangle&=&\ S(S+1)\ |\Psi\rangle\\
 S_z|\Psi\rangle&=&\left(\frac{N_{\uparrow}-N_{\downarrow}}{2}\right)|\Psi\rangle.
\eea
For ferromagnetic state $S=N/2$.

For the finite value of the layer separation $d/l$,
the direct part with zero wave vector contributes as a static charging energy
$(2d/N)S_z^2$ \cite{mac1} which makes $S_z=0$. Because the symmetry of the Hamiltonian
is reduced from SU(2) to U(1), the ground state
is no longer the eigen state of ${\bf S}^2$ although it still is an
 eigen state of $S_z$. 
We define the length of the total pseudospin by the expectation value
 $\langle\Psi|{\bf S}^2|\Psi\rangle=S(S+1)$.
In Fig.2 the value of $S/N$ as a function of $d/l$ is plotted.
The deviation
from the ferromagnetic state with increase of $d/l$ can be seen.

\begin{center}
\begin{figure}
\epsfxsize 74mm \epsffile{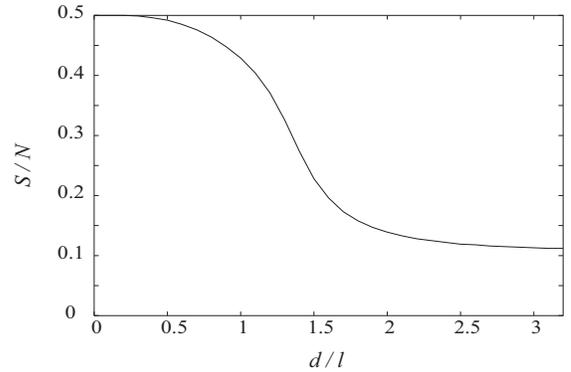}
\caption{The value of $S/N$, where $S$ is the length of the total pseudospin,
 is plotted  as a function of the 
layer spacing. The number of electron is $N=12$.
}
\label{}
\end{figure}
\end{center}

To investigate the nature of the ground state we calculate the two-particle
correlation function $g_{\sigma\sigma'}({\bf r})$ which is defined as
\bea
 g_{\sigma\sigma'}({\bf r})=
 \frac{1}{N(N-1)}\sum_{\bf q} {\rm e}^{{\rm i}{\bf q}\cdot{\bf r}}
\ S_{\sigma\sigma'}({\bf q})
\eea
where
\bea
S_{\sigma\sigma'}({\bf q})=\langle\Psi|\rho_{\sigma}(-{\bf q})
\rho_{\sigma'}({\bf q})|\Psi\rangle
\eea
is the structure factor.
 Fig.2(a) shows the $g_{\uparrow\downarrow}({\bf r})$
 (solid line) and $g_{\uparrow\uparrow}({\bf r})$ (dashed line) 
in the system with $d/l=0.3$. 
Around the origin,
both $g_{\uparrow\downarrow}({\bf r})$ and
 $g_{\uparrow\uparrow}({\bf r})$ behave as $r^2$.
$g_{\uparrow\downarrow}(0)=0$  shows a clear correlation hole.
 Such a behavior is the characteristic of
the $\Psi_{1,1,1}$ state. 
Actually overlap between the $\Psi_{1,1,1}$ and the wave function of the exact ground
state is almost unity in this regime.\cite{ymg}

Next we consider the case $d/l=0.9$.
This regime may correspond to that in which the Josephson-like effect was observed
in Spielman {\it {et al}}'s experiment.
Actually Fig.1 shows that the ferromagnetic order or the interlayer
phase coherence survives at this point.
The system should be in a quantum Hall regime
according to several experimental data\cite{murphy,spl} 
and theoretical analysis\cite{mac1,moon},
 but $\Psi_{1,1,1}$ is not relevant
at this point.\cite{ymg} The small but finite value of $g_{\uparrow\downarrow}(0)$
indicates that the correlation hole deflates gradually.
The decline of the correlation hole may lead to
non-singular behavior of the zero-bias conductivity.
The $g_{\uparrow\downarrow}({\bf r})$ in Fig.2(b) shows not only short range interlayer
correlation but the fact that
the electrons in the `down' layer is attracted to the electron in the `up' layer 
at the origin.

To understand this state we consider an effective model state which is defined as a 
ground state of the following pseudo potential interaction;\cite{qhe}
$ V^{\uparrow\uparrow}_l=0,
 V^{\uparrow\downarrow}_l=-\delta_{l,1}$.

\begin{center}
\begin{figure}
\epsfxsize 90mm \epsffile{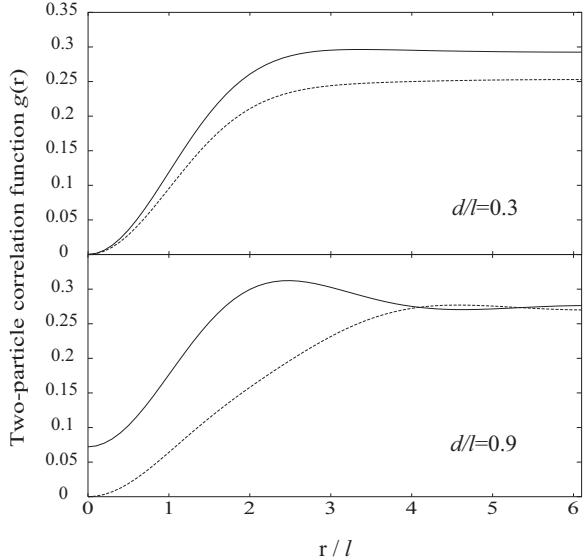}
\caption{
The two-particle correlation
 function $g_{\uparrow\downarrow}$ (solid line), $g_{\uparrow\uparrow}$ (dashed line)
at (a) $d/l=0.3$, and (b) $d/l=0.9$ are plotted. 
The total number of electron is $N$ = 12.
}
\label{}
\end{figure}
\end{center} 
\noindent
The negative sign in $V_l^{\uparrow\downarrow}$ means interlayer attraction
in the $l=1$ channel of the angular momentum.
This state has the imperfect correlation holes.
On the other hand,
 $\Psi_{1,1,1}$ is defined as the ground state of the pseudo potential interaction
$ V^{\uparrow\uparrow}_l=V^{\uparrow\downarrow}_l=\delta_{l,0}$.
In Fig.3, overlap between the exact ground state and these model states are shown.
We find our effective model becomes relevant at $d/l=0.85$.
The maximum value of overlap is $0.996$.

\begin{center}
\begin{figure}
\epsfxsize 85mm \epsffile{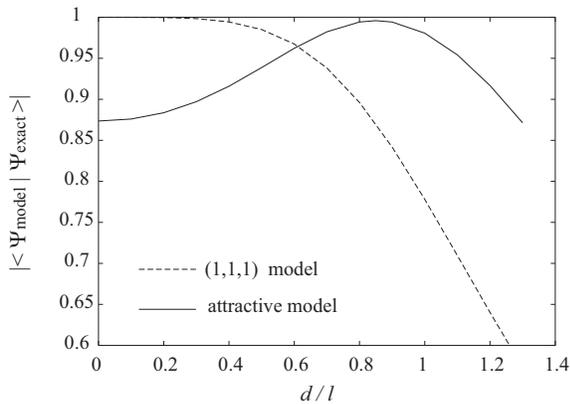}
\caption{
Overlap between the exact ground state 
and the model states as functions of $d/l$.
The number of electron is $N=12$.
}
\label{}
\end{figure}
\end{center} 


To see the nature of the crossover between Halperin's (1,1,1) model and our
attractive model, we consider a model Hamiltonian with pseudo potential interaction
$V_l^{\uparrow\downarrow}=\cos\theta\delta_{l,0}-\sin\theta\delta_{l,1}$,
 $V_l^{\uparrow\uparrow}=V_l^{\downarrow\downarrow}=0$.
We calculated the relevant value of $\theta$ which has a maximum overlap with 
the exact ground state as a function of $d/l$ and
found a continuum transition from $\theta=0$ ((1,1,1) model) to
 $\theta=\pi/2$ (attractive model). 
This is the decay of correlation hole.
When the layer spacing $d$ exceeds $1.4l$, the nature of  correlation changes
qualitatively.
The two-particle correlation functions $g_{\uparrow\uparrow}$ and
$g_{\uparrow\downarrow}$ exhibit a $2k_F$-like oscillation.
 Such a behavior was discussed in
a monolayer system with $\nu=1/2$ by Rezayi and Read\cite{rr} where
$k_F$ ($=1/l$ in $\nu=1/2$ \cite{hlr}) is a Fermi wave number
of the composite fermion.

Finally we consider the evolution of the ground state based on a modified 
composite fermion theory.
This theory of the bilayer system was formulated by Rajaraman.
\cite{rajaraman} A filling factor $\nu_{\rm T}=1$ system was discussed by  several
authors.\cite{bonesteel,kim,morinari,veillette}
We introduce composite fermion feild;
 $\phi_{\sigma}^{\dag}({\bf r})=\psi_{\sigma}^{\dag}({\bf r})\
{\rm e}^{J_{\sigma}({\bf r})}$
where $\psi_{\sigma}$ is the electrons field operator and
 $J_{\sigma}({\bf r})=R_{\sigma\sigma'}\int d^2{\bf r}' \rho_{\sigma'}({\bf r}')
{\rm ln}(z-z') -\frac{eB}{4}|{\bf r}|^2$ has a role of the flux attachment.\cite{read}
The matrix $R$ is appropriately determined by interlayer interaction
at a corresponding layer separation.
For $\nu_{\rm T}=1$ bilayer system, when the layer separation is sufficiently large, the Fermi-liquid state
of the composite fermions with $R_{\sigma\sigma'}=2\delta_{\sigma,\sigma'}$
 corresponds to eq.(2).
On the other hand, for small $d/l$, we should choose  
$R_{\sigma\sigma'}=2\delta_{\sigma,-\sigma'}$ because of strong interlayer interaction.
These theories are called independent composite fermion and mutual composite fermion
theories respectively.
The operation of the flux attachment is carried on by a two-body interaction that cause
a pairing instability of the composite fermions.\cite{kim,morinari,gww}
The wave function of the interlayer paired state is written as
\bea
 \Psi={\rm Pf}\ K({\bf r}_i^{\sigma}-{\bf r}_j^{-\sigma})\ 
\Psi_{R_{\uparrow\uparrow}, R_{\downarrow\downarrow},R_{\uparrow\downarrow}},
\label{pair}
\eea
where ${\rm Pf}K={\cal A}(K_{1,2}K_{3,4}\cdots)$ is an antisymmetric product and the factor
$
 K({\bf r}) = \frac{1}{\Omega}\sum_{\bf k} {\rm e}^{{\rm i}{\bf k}\cdot {\bf r}}
\left(\frac{ E_{\bf k}-\epsilon_{\bf k}}{\Delta_{\bf k}^*} \right)
\label{K}
$
is determined by the procedure of the Bogoliubov transformation.
$\Delta_{\bf k}$ is a pairing gap, $\epsilon_{\bf k}=k^2/2m^*-\epsilon_{\rm F}$, and
$E_{\bf k}=\sqrt{\epsilon_{\bf k}^2+|\Delta_{\bf k}|^2}$.
 ${\rm Pf}\ K({\bf r}_i^{\sigma}-{\bf r}_j^{-\sigma})$ is the wave function of the 
composite fermions that is a real space representation
of the BCS-like paired state.\cite{gww}
When $d/l$ is small, we consider the case $R_{\sigma,\sigma'}=2\delta_{\sigma,-\sigma'}$.
For a week pairing state in the terminology of Read and Green,\cite{rg}
in the $l_z=-1$ channel $\Delta_{\bf k}=\Delta\cdot\left(\frac{k_x-{\rm i}k_y}{k_F} \right)$,
$K({\bf r})=1/z$.
Using the Cauchy identity,\cite{hr} one obtain $\Psi_{1,1,1}$ from eq.(\ref{pair}). 
On the other hand, for large $d/l$ case, Kim {\it et al}. proposed several paired state
of the independent composite fermions.\cite{kim}
However 
anything of that kind was not found in our calculation.
It rather seems that our results 
of the two-particle correlation function
indicate the continuum development
of electrons pairing in $l_z=1$ channel that corresponds to $l_z=-1$ 
pairing of the mutual composite fermions.
This is understood as follows.
Note the interlayer Coulomb interaction reduces the interlayer pairing.
Starting from $d=0$, where the mutual composite fermion theory gives $\Psi_{1,1,1}$
as the ground state,
 increase of $d/l$ makes the interlayer pairing develop further
 because of the reduction of the interlayer interaction.
We note that the increase of $d/l$ also decreases a spin stiffness coefficient 
and makes the charge gap decrease.\cite{moon}
In Fig.4, the $r$-dependence of the function $|K(r)|$ with two case of 
the intensity of the interlayer pairing $\Delta/\epsilon_{\rm F}$ are plotted.
We found that
at $\Delta /\epsilon_{\rm F}=\sqrt{2}$, $|K(r)|$ coincides with $1/r$ completely
that lead to $\Psi_{1,1,1}$.
At $\Delta/\epsilon_{\rm F}=3$, $|K(r)|$ behaves as $1/r^{1.4}$ asymptoticaly
 at large $r$ $(>10 l)$
and $1/r^2$ at small $r$ $(<10 l)$.
This $1/r^2$ behavior at small $r$ with eq.(\ref{pair}) 
is consistent with the contraction of the correlation hole
 seen in Fig.2(b).
\begin{center}
\begin{figure}
\epsfxsize 80mm \epsffile{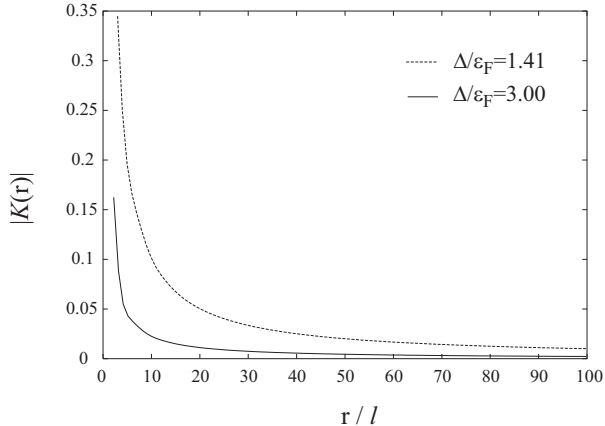}
\caption{
The $r$-dependence of the absolute value of $K({\bf r})$ in eq (\ref{pair}).
}
\label{}
\end{figure}
\end{center}

In this letter we have investigated the evolution of the bilayer quantum Hall
ferromagnetic state.
Around $d/l\sim 1$ $\Psi_{1,1,1}$ is not relevant as a trial wave function.
We proposed an attractive model that has imperfect correlation holes
and gives a good approximation to the ground state.
In our model state, the electrons are interacting attractively
in $l_z=1$ channel which is connected to the paired state of 
the mutual composite fermion theory.
The crossover corresponds to a decline of the correlation hole that
is consistent with the fact that as $d/l$ is increased the peak of the zero-bias
 conductance and the Hall drag resistance
 turned down in the experiments.\cite{spl,kellogg} 


We are grateful to N. Shibata and A. H. MacDonald
for useful
discussions and members of MacDonald group in the University of
 Texas for their hospitality.
This work is supported by a Grant-in-Aid for Scientific Research (C) 10640301
from the Ministry of Education, Science, Sports and Culture.
%
%
%

\end{multicols}

\end{document}